\documentclass[sigconf]{acmart}

\newif\ifshowcomments
\showcommentsfalse 

\ifshowcomments
    \newcommand{\TA}[1]{\textcolor{blue}{\textbf{*TA*}: #1}}

    \newcommand{\todo}[1]{\textcolor{red}{\textbf{(TODO)}: #1}}
\else
    \newcommand{\TA}[1]{}
    \newcommand{\DK}[1]{}
    \newcommand{\todo}[1]{}
    \newcommand{\nn}[1]{}
\fi

\AtBeginDocument{%
  }

\usepackage[frozencache]{minted}

\usepackage[table]{xcolor}

\setcopyright{acmlicensed}
\copyrightyear{2018}
\acmYear{2018}
\acmDOI{XXXXXXX.XXXXXXX}
\acmConference[Conference acronym 'XX]{Make sure to enter the correct
  conference title from your rights confirmation email}{June 03--05,
  2018}{Woodstock, NY}
\acmISBN{978-1-4503-XXXX-X/2018/06}

\begin{document}

\title{The Stochastic Shift: A New Evaluation Paradigm for Text-to-SQL with AI Operators}




\author{Tarfah Alrashed}
\affiliation{%
  \institution{Google}
  \city{}
  \country{}
}
\email{tarfah@google.com}

\author{Fatma \"Ozcan}
\affiliation{%
  \institution{Google}
  \city{}
  \country{}
  }
\email{fozcan@google.com}

\author{Per Jacobsson}
\affiliation{%
  \institution{Google}
  \city{}
  \country{}
  }
\email{pjacobsson@google.com}

\author{Tal Neiman}
\affiliation{%
  \institution{Google}
  \city{}
  \country{}
  }
\email{talneiman@google.com}

\author{Xianshun Chen}
\affiliation{%
  \institution{Google}
  \city{}
  \country{}
  }
\email{xianshun@google.com }

\renewcommand{\shortauthors}{Alrashed et al.}

\begin{abstract}

SQL has been augmented with AI operators, enabling modern data analytics platforms to derive insights from both structured and unstructured data. We observe that while current Text-to-SQL systems can successfully generate these AI-augmented queries, reliably evaluating their correctness remains a critical open challenge. Current metrics, which rely on exact query results and deterministic execution, systematically fail against the flexible, non-deterministic outputs of AI operators. In this paper, we formalize these unique evaluation failure modes and introduce a \emph{Multilayered Evaluation Framework} that decouples deterministic database logic from flexible AI semantics. We test our approach across both industry (BigQuery) and academic (ThalamusDB) systems. We demonstrate that traditional Execution Accuracy severely penalizes valid queries, achieving as low as a 25\% detection rate for correct translations. Furthermore, even a state-of-the-art LLM-based autorater falsely rejects 32\% of accurate queries due to the complexity of judging both relational and AI components simultaneously. By validating the standard relational logic and the AI operations separately, our framework achieves state-of-the-art overall accuracy across both platforms (up to 97.2\%), proposing a reliable standard for benchmarking AI-powered SQL generators.

\end{abstract}




\maketitle

\section{Introduction}

Modern database management systems are actively evolving beyond structured tabular data by embedding AI operators directly into SQL. Platforms like Google’s BigQuery, AlloyDB, Snowflake~\cite{liskowski2026cortex}, and ThalamusDB~\cite{10.1145/3654989} now provide native AI functions and vector search capabilities (e.g., \texttt{AI.IF}, \texttt{AI.SEARCH})~\cite{patel2026ainative,patel2024semanticoperators}. These operators allow a database to process unstructured data directly within a SQL query. For example, filtering a table of images to find specific content---such as querying \texttt{SELECT car\_id WHERE AI.IF("is this car red?", car\_image)}---bridges a semantic gap that was previously impossible in traditional relational databases. Because writing these complex, hybrid queries manually can hinder adoption, translating natural language into AI-augmented SQL has emerged as a critical next step.

However, this paradigm shift completely breaks current Text-to-SQL evaluation pipelines. Historically, the NLP and database communities optimized to translate natural language into precise, relational code~\cite{sun2024sqlpalmimprovedlargelanguage,1864519.1864543}. As a result, the highly brittle Exact Match accuracy~\cite{pourreza2023evaluating} gave way to Execution Accuracy (EX)~\cite{gan-etal-2021-natural-sql,bird} and Test Suite Accuracy (TSA)~\cite{zhong-etal-2020-semantic} to reduce false negatives. However, a core assumption underlies all these Text-to-SQL evaluation metrics: they were designed for an environment where SQL queries execute deterministically over structured, tabular data.

Unlike User-Defined Functions (UDFs) that execute mathematically sound, deterministic logic, LLMs rely on flexible, interpretive reasoning~\cite{patel2026ainative,khattab2023dspy}. Embedded AI functions require a shift away from strict mathematical equivalence and toward ``relaxed semantic correctness.'' When evaluated under current metrics, valid queries fail completely. Exact Match protocols fail immediately because an LLM might inject slightly different but functionally equivalent phrasing into an operator's prompt (e.g., ``is there a cat in this image?'' vs. ``does this picture contain a cat?''), instantly triggering a false negative. EX suffers a similar fate; because slightly different prompt phrasing can shift the LLM's decision boundary on subjective, borderline cases, executing these functionally equivalent prompt variations over unstructured data often yields minor, acceptable differences in the returned rows. In short, current metrics enforce rigid binary correctness---artificially penalizing valid generative reasoning and completely failing to interpret user intent in this hybrid domain.

To overcome the limitations of strict execution metrics, the research community has increasingly explored LLM-based autoraters to evaluate semantic correctness~\cite{zhao2025llm}. However, as we observe in our experiments, evaluating a query that combines deterministic database logic and probabilistic AI prompts overloads these autoraters. This creates an evaluation conflict: the LLM struggles to simultaneously verify strict relational boundaries (e.g., tracking \texttt{DISTINCT} or \texttt{UNION} constraints) while performing fuzzy semantic matching on the embedded AI prompts, leading them to falsely reject a large number of valid queries. Consequently, creating reliable benchmarks for Text-to-SQL with AI operators necessitates a new evaluation methodology that separates the deterministic relational part of a query from the flexible semantic evaluation of its AI prompts. In this paper, we expose the unique failure modes of AI-augmented SQL when evaluated via EX and a state-of-the-art autorater, and present a novel benchmarking architecture engineered to decouple relational execution from semantic validation.

Specifically, this paper makes the following contributions:
\begin{itemize}
    \item We introduce a \emph{Multilayered Evaluation Framework} that uses a \emph{SQL Split} module to decouple deterministic relational logic from non-deterministic AI-augmented parts, paired with an LLM \emph{autorater} to evaluate prompt intent independently of database execution.
    
    \item We demonstrate that while EX severely penalizes valid AI-augmented queries (detecting as few as 25\% of correct translations), and state-of-the-art LLM autoraters struggle with hybrid evaluation (falsely rejecting up to 32\% of accurate queries), our decoupled framework achieves state-of-the-art overall accuracy across both platforms (up to 97.2\%).
    
    \item We formally categorize the failure modes of Text-to-SQL generation with AI operators and outline a roadmap of open challenges for future generative query builders.

    \item  We open-source our complete evaluation framework, including all baseline evaluation scripts and prompting templates: \url{https://figshare.com/s/c75141469fb6dc74b890}
\end{itemize}

\section{Background}
\label{sec:background}

Standard SQL and relational databases were designed for structured, tabular data. Traditional operators rely on deterministic logic to filter, join, or aggregate numerical and categorical fields, but they cannot comprehend unstructured data like free-text, images, or audio. Historically, extracting insights from such data required exporting it out of the database into external machine learning pipelines, introducing significant data-movement overhead and architecture complexity.

The integration of LLMs has addressed this limitation by embedding semantic understanding directly into the database via AI operators. By bringing generative models to the data, these operators allow users to query unstructured data using the familiar, declarative SQL interface. While LLM inference introduces computational costs, AI operators bridge the gap between strict relational logic and probabilistic reasoning, enabling tasks like semantic filtering and classification without requiring external pipelines or task-specific fine-tuning.

This in-database approach is gaining rapid support in both academia and industry~\cite{lotus,sun2024sqlpalmimprovedlargelanguage}. Academic systems like ThalamusDB provide operators such as \texttt{NLfilter} and \texttt{NLjoin} to process multi-modal data, while enterprise platforms like Google Cloud's BigQuery now include built-in AI functions (e.g., \texttt{AI.IF}, \texttt{AI.SCORE}, and \texttt{AI.SEARCH}).

To illustrate, suppose a user wants to find electric cars with a dead battery based on diagnostic audio recordings. Traditional SQL cannot filter on raw audio content. However, with AI operators, this becomes a straightforward query.  
In BigQuery, an \texttt{AI.IF} operator evaluates a natural language prompt directly against the audio:
\begin{minted}[style=perldoc, fontsize=\small]{sql}
SELECT DISTINCT cars.car_id FROM cars, car_audio
WHERE cars.car_id = car_audio.car_id
  AND cars.fuel_type = 'Electric'
  AND AI.IF(("Return true if the car recording indicates 
  a dead battery, false otherwise.", car_audio.audio_path), 
  connection_id => '...');
\end{minted}

Academic systems offer analogous functions. For example, ThalamusDB executes the same intent using the \texttt{NLfilter} command:
\begin{minted}[style=perldoc, fontsize=\small]{sql}
SELECT DISTINCT cars.car_id FROM cars, car_audio
WHERE cars.car_id = car_audio.car_id
  AND cars.fuel_type = 'Electric'
  AND NLfilter(car_audio.audio_path, "Return true if the car 
  recording indicates a dead battery, false otherwise.");
\end{minted}

These examples show AI operators combining deterministic filtering on structured data (e.g., \texttt{fuel\_type = 'Electric'}) with stochastic, semantic evaluation of unstructured content.

\section{The Breakdown of Current Evaluation}

Current performance metrics systematically fail when evaluating Text-to-SQL with AI operators. While previous studies have already highlighted flaws in EX for standard SQL~\cite{zhong-etal-2020-semantic, pourreza2023evaluating}, the introduction of AI operators causes this metric to break down completely. To illustrate why existing pipelines break down, consider this natural language request: \textit{``Count of positive reviews for movie taken\_3. Return the count as positive\_review\_cnt.''}

A benchmark dataset might define the ``gold-standard'' reference query as follows:
\begin{minted}[style=perldoc, fontsize=\small]{sql}
SELECT COUNT(*) AS positive_review_cnt 
FROM movie.reviews 
WHERE id = 'taken_3' 
  AND AI.IF(('Determine if the following movie review is 
  positive:', reviewText), connection_id => '...');
-- Execution Result: 9
\end{minted}

Now suppose a Text-to-SQL model generates this functionally equivalent query:

\begin{minted}[style=perldoc, fontsize=\small]{sql}
SELECT COUNT(*) AS positive_review_cnt 
FROM movie.reviews 
WHERE id = 'taken_3' 
  AND AI.IF(('Is this review positive? Review:', reviewText),
  connection_id => '...');
-- Execution Result: 11
\end{minted}

Both queries are valid translations of the user's intent. In fact, analyzing underlying data reveals that the 11 reviews returned by the generated SQL explicitly include all 9 benchmark reviews. Yet, under EX, the primary evaluation metric for modern leaderboards, this functionally correct generation scores a 0\%. 

This failure occurs because EX assumes execution is strictly deterministic. While standard SQL filters (e.g., \texttt{id = `taken\_3'}) return absolute binary results, AI operators do not. Slightly different prompt phrasing could cause the underlying LLM to shift its decision boundary on subjective, borderline cases. Because the generated query categorized 11 reviews as positive while the benchmark categorized 9, EX enforces rigid mathematical equivalence (\texttt{11 != 9}) and falsely rejects the generated query.

This rigid rejection exposes a fundamental incompatibility: current benchmarks demand identical data intersections, whereas AI-augmented SQL relies on probabilistic semantic analysis. Evaluating this emerging domain requires a shift toward \textit{relaxed semantic correctness}~\cite{zhao2025llm}. The objective is no longer to achieve strict set equivalence, but to capture the user's intent. If a generated query accurately targets the semantic goal---even if the data varies slightly due to LLM sampling---it should be accepted as correct. Automating the measurement of this semantic alignment forms the theoretical foundation of our Multilayered Evaluation Framework.

\section{Multilayered Evaluation Framework}
\begin{figure*}
  \includegraphics[width=\textwidth]{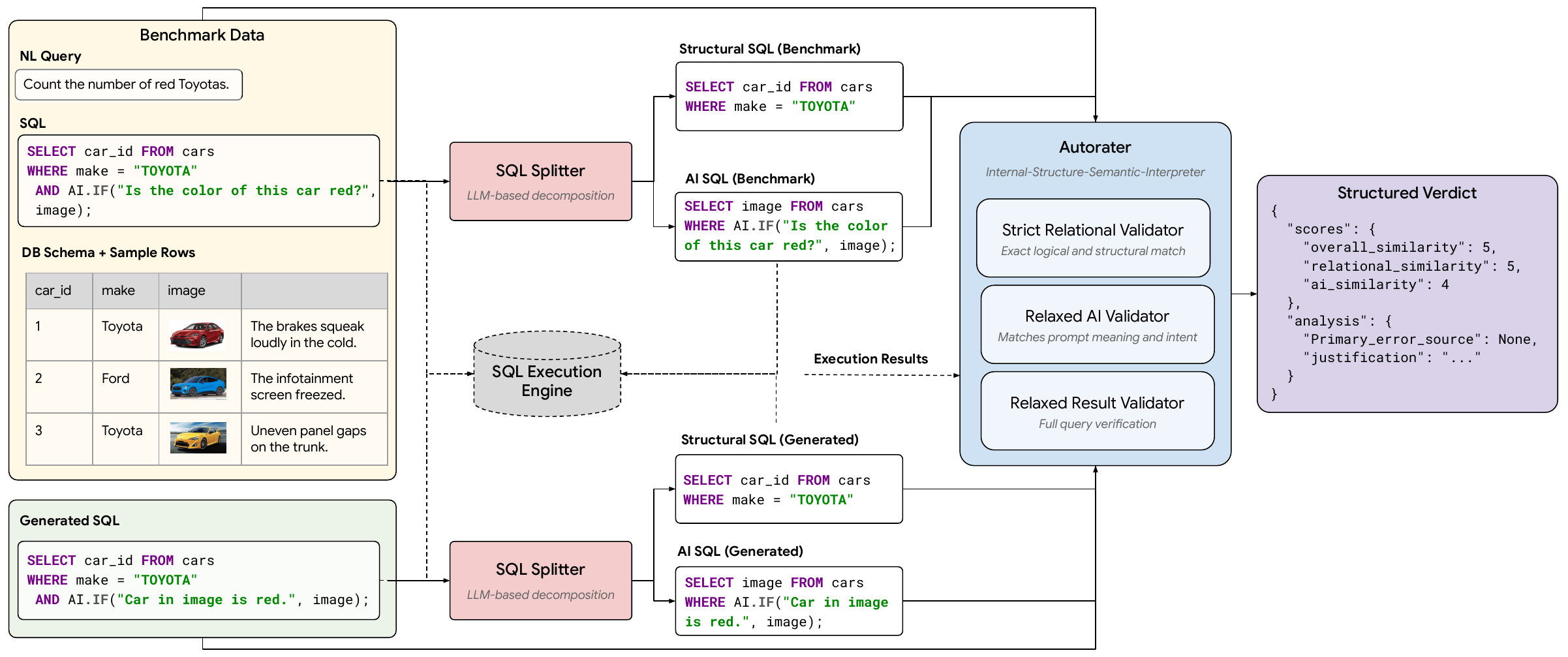}
  \caption{Overview of the Multilayered Evaluation Framework for AI-augmented SQL, operating in three phases: Splitting (decoupling relational structure from AI components), Execution (running queries and isolated components against the database), and Autorating (independent semantic evaluation via an LLM-as-a-judge).}
  \Description{The architecture of our Multilayered Evaluation Framework.}
  \label{fig:eval-framework}
\end{figure*}

To address the limitations of existing evaluation methods, we propose a Multilayered Evaluation Framework designed specifically for the dual nature of AI-augmented SQL. As illustrated in Figure~\ref{fig:eval-framework}, our framework operates in three phases:

\begin{enumerate}
    \item \textbf{Splitting:} Decoupling a query's deterministic relational structure from its AI component.
    \item \textbf{Execution:} Running both the full queries (benchmark and generated) and their newly isolated split components against the database to gather their respective result sets.
    \item \textbf{Autorating:} Feeding the database schema, natural language question, queries, and execution results into an LLM-as-a-judge for independent semantic evaluation.
\end{enumerate}

By isolating the standard SQL structure from the AI tasks and capturing their distinct execution behaviors, we prevent valid generative queries from being falsely rejected due to language variations, non-deterministic execution, or the use of alternative AI operators.

\paragraph{SQL Splitter}
Because current metrics strictly penalize any variation or non-determinism within the AI component, our framework first isolates the structural validity of a relational query from its semantic intent. The ``SQL Splitter'' is a general architectural component that can be implemented in various ways, such as using a traditional Abstract Syntax Tree (AST) parser or an LLM.

We use a few-shot LLM-based SQL Splitter. Given the relative simplicity of SemBench queries, it proved highly effective at dividing benchmark and generated queries into structural and AI components, as shown below:

\begin{minted}[style=perldoc, fontsize=\footnotesize]{sql}
-- Original SQL
SELECT p.car_id FROM cars_dataset.car_mm as x
JOIN cars_dataset.cars  AS p ON p.car_id = x.car_id
WHERE AI.IF(prompt => ("Returns true if the vehicle image 
contains both a puncture and paint scratches.", x.image), 
connection_id => '...') LIMIT 100;
\end{minted}

\begin{minted}[style=perldoc, fontsize=\footnotesize]{sql}
-- Split SQL (Structural Part)
SELECT p.car_id FROM cars_dataset.car_mm as x 
JOIN cars_dataset.cars AS p ON p.car_id = x.car_id 
WHERE TRUE LIMIT 100;
\end{minted}

\begin{minted}[style=perldoc, fontsize=\footnotesize]{sql}
-- Split SQL (AI Part)
SELECT x.image FROM cars_dataset.car_mm as x 
WHERE AI.IF(prompt => ("Returns true if the vehicle image 
contains both a puncture and paint scratches.", x.image), 
connection_id => '...') LIMIT 100;
\end{minted}

\paragraph{Independent Autorating}
Following the split, an LLM-based Autorater evaluates semantic similarity independently of strict execution matching. We feed the Autorater the natural language question, the database schema, the full queries (both benchmark and generated), their split components, and their respective execution results.  

The Autorater outputs a detailed breakdown of the query's accuracy. First, it scores overall, relational, and AI similarity on a 1 to 5 scale, as this 5-point grading has been shown to match human judgment best~\cite{li2026gradingscaleimpactllmasajudge}. We consider any score over 3 to be correct. This captures \textit{relaxed semantic correctness} by accepting functionally equivalent (4) or perfectly aligned (5) queries, while strictly rejecting ambiguous or partially correct outputs ($\le$3). Finally, the Autorater flags the primary source of any errors (e.g., relational logic vs. AI prompt text) and provides a clear, written explanation for its ratings.

\section{Experiments}

\subsection{Experimental Setup}

\paragraph{Model}
In our experiment, we used \emph{gemini-3.1-pro-preview} configured via the Google GenAI SDK (targeting the Google Cloud Vertex AI backend) using the default sampling parameters (temperature = 1.0, top-P = 0.95, and top-K = 40, with no random seed set). To generate executable SQL, the model relied on in-context learning with schema-specific instructions and programmatic few-shot examples. We enforced a structured output mode during SQL splitting and autorater evaluation to guarantee schema-conforming responses. Finally, the benchmark and generated queries were executed directly in their respective database engines (BigQuery and ThalamusDB) to get execution data for evaluation.

\paragraph{Dataset} 
We used SemBench~\cite{lao2026sembench}, a benchmark designed for testing database queries on datasets containing text, image, and audio data. SemBench consists of 55 queries across five real-world scenarios, covering tasks like filtering, joining, classifying, and ranking multimodal data. Although SemBench includes queries for several systems (including Python-based tools like LOTUS and Palimpzest), we focused exclusively on BigQuery and ThalamusDB because our work targets Text-to-SQL systems with AI operators.

\paragraph{Evaluation Baselines} 
To empirically expose the failure modes of current evaluation methods, we compare our proposed \emph{Multilayered Evaluation Framework} against two baselines: \emph{Execution Accuracy (EX)} and a state-of-the-art LLM metric called \emph{LLM-SQL-Solver}~\cite{zhao2025llm}. 

LLM-SQL-Solver explores whether an LLM can be used to determine the equivalence between two SQL queries under two defined notions: semantic equivalence and relaxed equivalence. It is an entirely prompt-based approach that relies purely on the query text and schema, without any access to a real SQL engine or live execution data. To assist the LLM in this reasoning process, the authors of LLM-SQL-Solver presented two specific prompting techniques to force the model to output its intermediate reasoning before arriving at a final classification:

\begin{itemize}
    \item Miniature \& Mull (Semantic Equivalence): This technique evaluates semantic equivalence by asking the LLM to generate a small, imaginary database. The LLM then mentally simulates executing both queries on this mock data to find counterexamples yielding different results.
    \item Explain \& Compare (Relaxed Equivalence): This technique evaluates relaxed equivalence by guiding the LLM to first step through and explain the underlying logic of each query. It then compares these textual explanations to see if the queries share the same pragmatic intention, ignoring minor structural differences.
\end{itemize}

For our evaluation, we used both the \emph{Miniature \& Mull} and \emph{Explain \& Compare} approaches as our LLM baselines, passing them custom few-shot examples tailored to our AI operators. To ensure a fair comparison and isolate the actual impact of our SQL Split method, we also evaluated augmented versions of these baselines that included the execution results for both the benchmark and generated queries. This removes any potential information gap, ensuring that both the LLM baselines and our proposed framework have access to the exact same data during evaluation.

\paragraph{Text-to-SQL Generation with AI Operators}
To produce the queries needed to test our evaluation methods, we implemented a Text-to-SQL generator. This generator translates a natural language question, a database schema, and a small random sample of rows into an executable query using in-context learning (ICL) with a few-shot prompt. Based on preliminary experiments, we found this prompting approach was strictly necessary to teach the LLM the novel syntax required to write valid AI-augmented SQL. To establish a reliable baseline, we manually verified the generated outputs. This analysis revealed that our generator successfully produced correct translations for 70\% of the BigQuery queries and 80\% of the ThalamusDB queries. This variation created a realistic mix of correct and flawed queries to test our evaluation framework.

\subsection{Experimental Results}

While our manual verification confirmed good generation quality (70\% to 80\% success depending on the database platform), current automated evaluation methods scored this much lower across both systems (Table \ref{tab:evaluation-methodologies}).


EX failed drastically on both platforms. It correctly identified only 36.0\% of valid BigQuery translations and just 25.0\% for ThalamusDB, leading to terrible overall accuracies (52.8\% and 26.7\%). It also failed to consistently reject flawed queries. This highlights a critical flaw in relying solely on execution results: in one case, a generated query left out a required \texttt{DISTINCT} clause, but because the benchmark database currently had no duplicate values, both queries returned the same outcome. EX incorrectly accepted this flawed query.

\begin{table*}[htbp]
\centering
\caption{Performance comparison of evaluation methods against our proposed framework on SemBench across two AI-augmented database systems (BigQuery and ThalamusDB). Rates reflect the correct acceptance of valid queries (True Positive) and correct rejection of flawed queries (True Negative).}
\label{tab:evaluation-methodologies}
\begin{tabular}{l|ccc|ccc}
\toprule
& \multicolumn{3}{c|}{\textbf{BigQuery}} & \multicolumn{3}{c}{\textbf{ThalamusDB}} \\
\cmidrule(lr){2-4} \cmidrule(lr){5-7}
\textbf{Evaluation Methodology} & \textbf{TPR} & \textbf{TNR} & \textbf{Accuracy} & \textbf{TPR} & \textbf{TNR} & \textbf{Accuracy} \\
\midrule
\textbf{Execution Accuracy (EX)} & 36.0\% & 90.9\% & 52.8\% & 25.0\% & 33.3\% & 26.7\% \\
\textbf{LLM-SQL-Solver (Miniature \& Mull)} & 68.0\% & \textbf{100.0\%} & 77.8\% & 66.7\% & \textbf{100.0\%} & 73.3\% \\
\textbf{LLM-SQL-Solver (Miniature \& Mull w/ Exec Results)} & 56.0\% & 90.0\% & 66.7\% & 58.3\% & \textbf{100.0\%} & 66.7\% \\
\textbf{LLM-SQL-Solver (Explain \& Compare)} & 76.0\% & 90.9\% & 80.6\% & \textbf{91.7\%} & 66.7\% & 86.7\% \\
\textbf{LLM-SQL-Solver (Explain \& Compare w/ Exec Results)} & 84.0\% & 81.8\% & 83.3\% & \textbf{91.7\%} & 66.7\% & 86.7\% \\
\textbf{Multilayered Eval Framework (Ours)} & \textbf{100.0\%} & 90.9\% & \textbf{97.2\%} & \textbf{91.7\%} & \textbf{100.0\%} & \textbf{93.3\%} \\
\bottomrule
\end{tabular}
\end{table*}

The standard LLM-SQL-Solver (\emph{Explain \& Compare}) performed somewhat better by focusing on logic. However, evaluating the entire hybrid query at once proved difficult. It struggled to balance correct acceptance and rejection across systems---for instance, while it achieved a high 91.7\% True Positive Rate on ThalamusDB, it incorrectly accepted many flawed queries, dropping its True Negative Rate to just 66.7\%. Adding execution data to this baseline did not fix the problem. For example, it incorrectly accepted a query that used \texttt{UNION DISTINCT} instead of the required \texttt{UNION ALL} simply because both operations happened to return the same results on this specific database.

We also evaluated the \emph{Miniature \& Mull} variation, which asks the LLM to mentally generate a counterexample database. Because it generated its own mock data, the LLM successfully imagined a database with duplicate values, allowing it to easily catch the missing \texttt{DISTINCT} error that fooled EX (achieving a perfect 100.0\% True Negative Rate on both systems). However, this counterexample generation falsely rejected around 32\% of perfectly correct queries across both databases. Furthermore, when we added execution results to this baseline, the real data directly conflicted with the prompt's command to invent counterexamples, causing its overall accuracy to drop even further.

Our Multilayered Evaluation Framework solved these issues by separating the relational logic from the AI component. This allowed it to safely use execution data without being tricked by the execution results. By evaluating the parts separately, it correctly caught the \texttt{UNION} error that fooled the EX baseline. It consistently provided the best balance, capturing 100\% of correct translations on BigQuery and achieving a perfect 100\% True Negative Rate on ThalamusDB, ultimately setting a state-of-the-art overall accuracy for both platforms (97.2\% and 93.3\% respectively).

\section{Open Challenges in Text-to-SQL Generation with AI Operators}
\label{sec:aisql-challenges}

The integration of AI operators unlocks powerful semantic capabilities, but it simultaneously introduces new challenges in query generation. While our Multilayered Evaluation Framework establishes a reliable methodology for evaluating these non-deterministic queries, generating them correctly presents failure modes distinct from traditional Text-to-SQL. Based on our preliminary analysis, we outline four critical challenges that future models must resolve.

\paragraph{Structured vs. Semantic Misalignment} 
The fundamental assumption in query routing is that deterministic execution over structured data is cheaper, faster, and more consistent than invoking AI operators. Meanwhile, embedding-based vector search occupies a middle ground—slower and less precise than relational logic, but much faster than AI operator inference. Generative agents frequently struggle to identify the optimal threshold along this spectrum. For instance, given the prompt \textit{``Show me rentals that include laundry,''} an agent must determine if a strict check against a boolean \texttt{has\_laundry} column is sufficient, or if it must compute an \texttt{AI.IF} function over an unstructured description column. Conversely, models often over-rely on structured data, such as mapping \textit{``three bedrooms''} to a deterministic filter \texttt{number\_of\_rooms = 3}, which erroneously flags houses with three total rooms instead of parsing the unstructured property description.

\paragraph{Data Source and Modality Ambiguity}
Subjective queries require the generative agent to selectively route logic to the most viable unstructured data source. If a user asks to \textit{``Show me pet-friendly rentals,''} the agent must independently decide whether the most accurate signal lies in raw property descriptions, user review text, or an image analysis of the property photos. Composing an execution path---determining whether to process these modalities via sequential AI filters or synthesize them in a single multi-modal prompt---requires a new form of semantic query optimization.

\paragraph{The Open-World vs. Closed-World Divide}
Complex queries often cross the boundary between closed-world data (database columns) and open-world knowledge (external facts held by the LLM). 
A prompt demanding \textit{``Sedans under 20,000 with a strong reputation for winter reliability''} requires an intelligent split. The model must apply standard SQL filters for exact matches (\texttt{price < 20000 AND type = 'sedan'}) alongside an AI function (\texttt{AI.IF}) to evaluate subjective knowledge like vehicle reliability records.

\paragraph{Forced Answering of Ambiguous Queries}
A common flaw in LLMs is the ``forced answering problem,'' where models default to generating SQL even if the user's request is invalid or incomplete~\cite{lee2024trustsql}. For AI-augmented SQL, this behavior is computationally expensive. If a database lacks the data to answer an ambiguous question, allowing the LLM to generate a complex AI query---rather than simply refusing the prompt---wastes tokens and severely increases execution latency. Developing mechanisms to reject unanswerable queries, such as those proposed for standard Text-to-SQL in related work~\cite{lee2024trustsql}, before they trigger AI operators is a critical requirement for real-world systems.

\section{Related Work}

\paragraph{LLM-Powered Data Systems and AI-Augmented SQL}
The research community is increasingly developing LLM-powered data processing systems that merge the reasoning capabilities of foundation models with the robust execution engines of traditional databases. Recent systems include LOTUS from Stanford~\cite{lotus}, Palimpzest from MIT~\cite{liu2025palimpzest}, DocETL from UC Berkeley~\cite{10.14778/3746405.3746426}, and ThalamusDB from Cornell~\cite{10.1145/3654989}. Commercial products like BigQuery, AlloyDB, Snowflake, and Databricks already support these operations by augmenting standard SQL with AI operators.
To drive the adoption of these AI queries by agents and human analysts, a robust natural language interface is required. However, while translating natural language into AI-augmented SQL unlocks powerful analytical capabilities, it introduces fundamental evaluation challenges that existing Text-to-SQL evaluation methods are unequipped to handle, motivating our evaluation framework.

\paragraph{Evolution of Text-to-SQL Benchmarks} 
While current benchmarks like Spider~\cite{yu-etal-2018-spider} and BIRD~\cite{bird} drove traditional Text-to-SQL progress, evaluating AI-augmented systems requires specialized datasets and relaxed metrics. Recently, SemBench~\cite{lao2026sembench} was proposed to assess LLM-powered multi-modal data systems. However, it primarily evaluates the computational cost, latency, and accuracy of the database systems themselves. To train and evaluate generative models for the natural language translation task, future benchmarks must contain a large volume of complex AI-operator queries. Critically, these datasets must incorporate mechanisms to measure relaxed semantic correctness. Only by moving beyond the rigid constraints of EX can the community accurately evaluate next-generation AI-powered query engines.

\paragraph{LLM-as-a-Judge and Semantic Evaluation}

To overcome the brittleness of execution-based metrics, recent natural language generation research has increasingly adopted LLM-as-a-Judge techniques~\cite{zhao2025llm}. These paradigms leverage foundation models to evaluate the semantic alignment and intention of generative outputs, rather than demanding execution equivalence. Our Multilayered Evaluation Framework extends this concept into the structured query domain by employing a SQL Split technique, isolating the deterministic relational logic from the stochastic AI part, and employing an autorater specifically tuned to assess the semantic intent of embedded AI operators.



\section{Conclusion \& Future Work}

To overcome the failure of strict execution metrics and baseline LLMs on AI-augmented SQL, we introduced a Multilayered Evaluation Framework that separates deterministic relational logic from non-deterministic AI components. Combining database execution with an independent LLM autorater, our framework achieves state-of-the-art accuracy on BigQuery (97.2\%) and ThalamusDB (93.3\%). Crucially, it reveals that current generators produce valid queries 70–80\% of the time—a success rate obscured by EX (which detected as few as 25\%) and existing LLM autoraters (which falsely rejected up to 32\% of valid outputs).
With this evaluation framework in place, our future work focuses on building a large-scale, challenging benchmark for Text-to-SQL with AI operators, targeting the failure modes in Section~\ref{sec:aisql-challenges}. By evaluating intelligent data routing, multi-modal ambiguity, and safe refusal, this dataset will serve as the definitive testing ground for next-generation query engines.




\bibliographystyle{ACM-Reference-Format}
\bibliography{sample-base}


\begin{thebibliography}{19}


\ifx \showCODEN    \undefined \def \showCODEN     #1{\unskip}     \fi
\ifx \showISBNx    \undefined \def \showISBNx     #1{\unskip}     \fi
\ifx \showISBNxiii \undefined \def \showISBNxiii  #1{\unskip}     \fi
\ifx \showISSN     \undefined \def \showISSN      #1{\unskip}     \fi
\ifx \showLCCN     \undefined \def \showLCCN      #1{\unskip}     \fi
\ifx \shownote     \undefined \def \shownote      #1{#1}          \fi
\ifx \showarticletitle \undefined \def \showarticletitle #1{#1}   \fi
\ifx \showURL      \undefined \def \showURL       {\relax}        \fi
\providecommand\bibfield[2]{#2}
\providecommand\bibinfo[2]{#2}
\providecommand\natexlab[1]{#1}
\providecommand\showeprint[2][]{arXiv:#2}

\bibitem[Gan et~al\mbox{.}(2021)]%
        {gan-etal-2021-natural-sql}
\bibfield{author}{\bibinfo{person}{Yujian Gan}, \bibinfo{person}{Xinyun Chen}, \bibinfo{person}{Jinxia Xie}, \bibinfo{person}{Matthew Purver}, \bibinfo{person}{John~R. Woodward}, \bibinfo{person}{John Drake}, {and} \bibinfo{person}{Qiaofu Zhang}.} \bibinfo{year}{2021}\natexlab{}.
\newblock \showarticletitle{{Natural {SQL}: Making {SQL} Easier to Infer from Natural Language Specifications}}. In \bibinfo{booktitle}{\emph{{Findings of the Association for Computational Linguistics: EMNLP 2021}}}, \bibfield{editor}{\bibinfo{person}{Marie-Francine Moens}, \bibinfo{person}{Xuanjing Huang}, \bibinfo{person}{Lucia Specia}, {and} \bibinfo{person}{Scott Wen-tau Yih}} (Eds.). \bibinfo{publisher}{{Association for Computational Linguistics}}, \bibinfo{address}{Punta Cana, Dominican Republic}, \bibinfo{pages}{2030--2042}.
\newblock
\href{https://doi.org/10.18653/v1/2021.findings-emnlp.174}{doi:\nolinkurl{10.18653/v1/2021.findings-emnlp.174}}


\bibitem[Jo and Trummer(2024)]%
        {10.1145/3654989}
\bibfield{author}{\bibinfo{person}{Saehan Jo} {and} \bibinfo{person}{Immanuel Trummer}.} \bibinfo{year}{2024}\natexlab{}.
\newblock \showarticletitle{{ThalamusDB: Approximate Query Processing on Multi-Modal Data}}.
\newblock \bibinfo{journal}{\emph{Proc. ACM Manag. Data}} \bibinfo{volume}{2}, \bibinfo{number}{3}, Article \bibinfo{articleno}{186} (\bibinfo{date}{May} \bibinfo{year}{2024}), \bibinfo{numpages}{26}~pages.
\newblock
\href{https://doi.org/10.1145/3654989}{doi:\nolinkurl{10.1145/3654989}}


\bibitem[Khattab et~al\mbox{.}(2023)]%
        {khattab2023dspy}
\bibfield{author}{\bibinfo{person}{Omar Khattab}, \bibinfo{person}{Arnav Singhvi}, \bibinfo{person}{Paridhi Maheshwari}, \bibinfo{person}{Zhiyuan Zhang}, \bibinfo{person}{Keshav Santhanam}, \bibinfo{person}{Sri Vardhamanan}, \bibinfo{person}{Saiful Haq}, \bibinfo{person}{Ashutosh Sharma}, \bibinfo{person}{Thomas~T Joshi}, \bibinfo{person}{Hanna Moazam}, {et~al\mbox{.}}} \bibinfo{year}{2023}\natexlab{}.
\newblock \showarticletitle{{DSPy: Compiling Declarative Language Model Calls into Self-Improving Pipelines}}.
\newblock \bibinfo{journal}{\emph{arXiv preprint arXiv:2310.03714}} (\bibinfo{year}{2023}).
\newblock


\bibitem[Lao et~al\mbox{.}(2026)]%
        {lao2026sembench}
\bibfield{author}{\bibinfo{person}{Jiale Lao}, \bibinfo{person}{Andreas Zimmerer}, \bibinfo{person}{Olga Ovcharenko}, \bibinfo{person}{Tianji Cong}, \bibinfo{person}{Matthew Russo}, \bibinfo{person}{Gerardo Vitagliano}, \bibinfo{person}{Michael Cochez}, \bibinfo{person}{Fatma Özcan}, \bibinfo{person}{Gautam Gupta}, \bibinfo{person}{Thibaud Hottelier}, \bibinfo{person}{H.~V. Jagadish}, \bibinfo{person}{Kris Kissel}, \bibinfo{person}{Sebastian Schelter}, \bibinfo{person}{Andreas Kipf}, {and} \bibinfo{person}{Immanuel Trummer}.} \bibinfo{year}{2026}\natexlab{}.
\newblock \bibinfo{title}{{SemBench: A Benchmark for Semantic Query Processing Engines}}.
\newblock
\showeprint[arxiv]{2511.01716}~[cs.DB]
\urldef\tempurl%
\url{https://arxiv.org/abs/2511.01716}
\showURL{%
\tempurl}


\bibitem[Lee et~al\mbox{.}(2024)]%
        {lee2024trustsql}
\bibfield{author}{\bibinfo{person}{Gyubok Lee}, \bibinfo{person}{Woosog Chay}, \bibinfo{person}{Seonhee Cho}, {and} \bibinfo{person}{Edward Choi}.} \bibinfo{year}{2024}\natexlab{}.
\newblock \bibinfo{title}{{TrustSQL: Benchmarking Text-to-SQL Reliability with Penalty-Based Scoring}}.
\newblock
\showeprint[arxiv]{2403.15879}~[cs.AI]
\urldef\tempurl%
\url{https://arxiv.org/abs/2403.15879}
\showURL{%
\tempurl}


\bibitem[Li et~al\mbox{.}(2023)]%
        {bird}
\bibfield{author}{\bibinfo{person}{Jinyang Li}, \bibinfo{person}{Binyuan Hui}, \bibinfo{person}{Ge Qu}, \bibinfo{person}{Jiaxi Yang}, \bibinfo{person}{Binhua Li}, \bibinfo{person}{Bowen Li}, \bibinfo{person}{Bailin Wang}, \bibinfo{person}{Bowen Qin}, \bibinfo{person}{Ruiying Geng}, \bibinfo{person}{Nan Huo}, \bibinfo{person}{Xuanhe Zhou}, \bibinfo{person}{Chenhao Ma}, \bibinfo{person}{Guoliang Li}, \bibinfo{person}{Kevin~C.C. Chang}, \bibinfo{person}{Fei Huang}, \bibinfo{person}{Reynold Cheng}, {and} \bibinfo{person}{Yongbin Li}.} \bibinfo{year}{2023}\natexlab{}.
\newblock \showarticletitle{{Can LLM already serve as a database interface? a big bench for large-scale database grounded text-to-SQLs}}. In \bibinfo{booktitle}{\emph{Proceedings of the 37th International Conference on Neural Information Processing Systems}} (New Orleans, LA, USA) \emph{(\bibinfo{series}{{NeurIPS} '23})}. \bibinfo{publisher}{Curran Associates Inc.}, \bibinfo{address}{Red Hook, NY, USA}, Article \bibinfo{articleno}{1835}, \bibinfo{numpages}{28}~pages.
\newblock


\bibitem[Li et~al\mbox{.}(2026)]%
        {li2026gradingscaleimpactllmasajudge}
\bibfield{author}{\bibinfo{person}{Weiyue Li}, \bibinfo{person}{Minda Zhao}, \bibinfo{person}{Weixuan Dong}, \bibinfo{person}{Jiahui Cai}, \bibinfo{person}{Yuze Wei}, \bibinfo{person}{Michael Pocress}, \bibinfo{person}{Yi Li}, \bibinfo{person}{Wanyan Yuan}, \bibinfo{person}{Xiaoyue Wang}, \bibinfo{person}{Ruoyu Hou}, \bibinfo{person}{Kaiyuan Lou}, \bibinfo{person}{Wenqi Zeng}, \bibinfo{person}{Yutong Yang}, \bibinfo{person}{Yilun Du}, {and} \bibinfo{person}{Mengyu Wang}.} \bibinfo{year}{2026}\natexlab{}.
\newblock \bibinfo{title}{{Grading Scale Impact on LLM-as-a-Judge: Human-LLM Alignment Is Highest on 0-5 Grading Scale}}.
\newblock
\showeprint[arxiv]{2601.03444}~[cs.CL]
\urldef\tempurl%
\url{https://arxiv.org/abs/2601.03444}
\showURL{%
\tempurl}


\bibitem[Liskowski et~al\mbox{.}(2026)]%
        {liskowski2026cortex}
\bibfield{author}{\bibinfo{person}{Pawe{\l} Liskowski}, \bibinfo{person}{Benjamin Han}, \bibinfo{person}{Paritosh Aggarwal}, \bibinfo{person}{Bowei Chen}, \bibinfo{person}{Boxin Jiang}, \bibinfo{person}{Nitish Jindal}, \bibinfo{person}{Zihan Li}, \bibinfo{person}{Aaron Lin}, \bibinfo{person}{Kyle Schmaus}, \bibinfo{person}{Jay Tayade}, {et~al\mbox{.}}} \bibinfo{year}{2026}\natexlab{}.
\newblock \showarticletitle{{Cortex AISQL: A Production SQL Engine for Unstructured Data }}. In \bibinfo{booktitle}{\emph{{Companion of the International Conference on Management of Data}}}. \bibinfo{pages}{400--412}.
\newblock


\bibitem[Liu et~al\mbox{.}(2025)]%
        {liu2025palimpzest}
\bibfield{author}{\bibinfo{person}{Chunwei Liu}, \bibinfo{person}{Matthew Russo}, \bibinfo{person}{Michael Cafarella}, \bibinfo{person}{Lei Cao}, \bibinfo{person}{Peter~Baile Chen}, \bibinfo{person}{Zui Chen}, \bibinfo{person}{Michael Franklin}, \bibinfo{person}{Tim Kraska}, \bibinfo{person}{Samuel Madden}, \bibinfo{person}{Rana Shahout}, {et~al\mbox{.}}} \bibinfo{year}{2025}\natexlab{}.
\newblock \showarticletitle{{Palimpzest: Optimizing ai-powered analytics with declarative query processing}}. In \bibinfo{booktitle}{\emph{{Proceedings of the Conference on Innovative Database Research (CIDR)}}}. \bibinfo{pages}{2}.
\newblock


\bibitem[Patel et~al\mbox{.}(2026)]%
        {patel2026ainative}
\bibfield{author}{\bibinfo{person}{Liana Patel}, \bibinfo{person}{Carlos Guestrin}, {and} \bibinfo{person}{Matei Zaharia}.} \bibinfo{year}{2026}\natexlab{}.
\newblock \showarticletitle{Towards AI-Native Data Systems with the Semantic Operator Model and LOTUS}.
\newblock \bibinfo{journal}{\emph{IEEE Data Engineering Bulletin}} (\bibinfo{year}{2026}).
\newblock
\urldef\tempurl%
\url{http://sites.computer.org/debull/A26mar/A26MAR-CD.pdf#page=61}
\showURL{%
\tempurl}


\bibitem[Patel et~al\mbox{.}(2024)]%
        {patel2024semanticoperators}
\bibfield{author}{\bibinfo{person}{Liana Patel}, \bibinfo{person}{Siddharth Jha}, \bibinfo{person}{Parth Asawa}, \bibinfo{person}{Melissa Pan}, \bibinfo{person}{Carlos Guestrin}, {and} \bibinfo{person}{Matei Zaharia}.} \bibinfo{year}{2024}\natexlab{}.
\newblock \showarticletitle{Semantic Operators: A Declarative Model for Rich, AI-based Analytics Over Text Data}.
\newblock  (\bibinfo{year}{2024}).
\newblock
\showeprint{2407.11418}
\urldef\tempurl%
\url{https://arxiv.org/abs/2407.11418}
\showURL{%
\tempurl}


\bibitem[Patel et~al\mbox{.}(2025)]%
        {lotus}
\bibfield{author}{\bibinfo{person}{Liana Patel}, \bibinfo{person}{Siddharth Jha}, \bibinfo{person}{Melissa Pan}, \bibinfo{person}{Harshit Gupta}, \bibinfo{person}{Parth Asawa}, \bibinfo{person}{Carlos Guestrin}, {and} \bibinfo{person}{Matei Zaharia}.} \bibinfo{year}{2025}\natexlab{}.
\newblock \bibinfo{title}{{Semantic Operators: A Declarative Model for Rich, AI-based Data Processing}}.
\newblock
\showeprint[arxiv]{2407.11418}~[cs.DB]
\urldef\tempurl%
\url{https://arxiv.org/abs/2407.11418}
\showURL{%
\tempurl}


\bibitem[Pourreza and Rafiei(2023)]%
        {pourreza2023evaluating}
\bibfield{author}{\bibinfo{person}{Mohammadreza Pourreza} {and} \bibinfo{person}{Davood Rafiei}.} \bibinfo{year}{2023}\natexlab{}.
\newblock \showarticletitle{{Evaluating cross-domain text-to-SQL models and benchmarks}}. In \bibinfo{booktitle}{\emph{{Proceedings of the 2023 Conference on Empirical Methods in Natural Language Processing}}}. \bibinfo{pages}{1601--1611}.
\newblock


\bibitem[Shankar et~al\mbox{.}(2025)]%
        {10.14778/3746405.3746426}
\bibfield{author}{\bibinfo{person}{Shreya Shankar}, \bibinfo{person}{Tristan Chambers}, \bibinfo{person}{Tarak Shah}, \bibinfo{person}{Aditya~G. Parameswaran}, {and} \bibinfo{person}{Eugene Wu}.} \bibinfo{year}{2025}\natexlab{}.
\newblock \showarticletitle{{DocETL: Agentic Query Rewriting and Evaluation for Complex Document Processing}}.
\newblock \bibinfo{journal}{\emph{{Proc. VLDB Endow.}}} \bibinfo{volume}{18}, \bibinfo{number}{9} (\bibinfo{date}{May} \bibinfo{year}{2025}), \bibinfo{pages}{3035–3048}.
\newblock
\showISSN{2150-8097}
\href{https://doi.org/10.14778/3746405.3746426}{doi:\nolinkurl{10.14778/3746405.3746426}}


\bibitem[Sun et~al\mbox{.}(2024)]%
        {sun2024sqlpalmimprovedlargelanguage}
\bibfield{author}{\bibinfo{person}{Ruoxi Sun}, \bibinfo{person}{Sercan~Ö. Arik}, \bibinfo{person}{Alex Muzio}, \bibinfo{person}{Lesly Miculicich}, \bibinfo{person}{Satya Gundabathula}, \bibinfo{person}{Pengcheng Yin}, \bibinfo{person}{Hanjun Dai}, \bibinfo{person}{Hootan Nakhost}, \bibinfo{person}{Rajarishi Sinha}, \bibinfo{person}{Zifeng Wang}, {and} \bibinfo{person}{Tomas Pfister}.} \bibinfo{year}{2024}\natexlab{}.
\newblock \bibinfo{title}{SQL-PaLM: Improved Large Language Model Adaptation for Text-to-SQL (extended)}.
\newblock
\showeprint[arxiv]{2306.00739}~[cs.CL]
\urldef\tempurl%
\url{https://arxiv.org/abs/2306.00739}
\showURL{%
\tempurl}


\bibitem[Yu et~al\mbox{.}(2018)]%
        {yu-etal-2018-spider}
\bibfield{author}{\bibinfo{person}{Tao Yu}, \bibinfo{person}{Rui Zhang}, \bibinfo{person}{Kai Yang}, \bibinfo{person}{Michihiro Yasunaga}, \bibinfo{person}{Dongxu Wang}, \bibinfo{person}{Zifan Li}, \bibinfo{person}{James Ma}, \bibinfo{person}{Irene Li}, \bibinfo{person}{Qingning Yao}, \bibinfo{person}{Shanelle Roman}, \bibinfo{person}{Zilin Zhang}, {and} \bibinfo{person}{Dragomir Radev}.} \bibinfo{year}{2018}\natexlab{}.
\newblock \showarticletitle{{S}pider: A Large-Scale Human-Labeled Dataset for Complex and Cross-Domain Semantic Parsing and Text-to-{SQL} Task}. In \bibinfo{booktitle}{\emph{Proceedings of the 2018 Conference on Empirical Methods in Natural Language Processing}}, \bibfield{editor}{\bibinfo{person}{Ellen Riloff}, \bibinfo{person}{David Chiang}, \bibinfo{person}{Julia Hockenmaier}, {and} \bibinfo{person}{Jun{'}ichi Tsujii}} (Eds.). \bibinfo{publisher}{Association for Computational Linguistics}, \bibinfo{address}{Brussels, Belgium}, \bibinfo{pages}{3911--3921}.
\newblock
\href{https://doi.org/10.18653/v1/D18-1425}{doi:\nolinkurl{10.18653/v1/D18-1425}}


\bibitem[Zelle and Mooney(1996)]%
        {1864519.1864543}
\bibfield{author}{\bibinfo{person}{John~M. Zelle} {and} \bibinfo{person}{Raymond~J. Mooney}.} \bibinfo{year}{1996}\natexlab{}.
\newblock \showarticletitle{{Learning to parse database queries using inductive logic programming}}. In \bibinfo{booktitle}{\emph{{Proceedings of the Thirteenth National Conference on Artificial Intelligence - Volume 2}}} (Portland, Oregon) \emph{(\bibinfo{series}{AAAI'96})}. \bibinfo{publisher}{{AAAI Press}}, \bibinfo{pages}{1050–1055}.
\newblock
\showISBNx{026251091X}


\bibitem[Zhao et~al\mbox{.}(2025)]%
        {zhao2025llm}
\bibfield{author}{\bibinfo{person}{Fuheng Zhao}, \bibinfo{person}{Jiayue Chen}, \bibinfo{person}{Lawrence Lim}, \bibinfo{person}{Ishtiyaque Ahmad}, \bibinfo{person}{Divyakant Agrawal}, {and} \bibinfo{person}{Amr El~Abbadi}.} \bibinfo{year}{2025}\natexlab{}.
\newblock \showarticletitle{{LLM-SQL-Solver: Can LLMs Determine SQL Equivalence?}}. In \bibinfo{booktitle}{\emph{{2025 IEEE International Conference on Big Data (BigData)}}}. IEEE, \bibinfo{pages}{1887--1894}.
\newblock


\bibitem[Zhong et~al\mbox{.}(2020)]%
        {zhong-etal-2020-semantic}
\bibfield{author}{\bibinfo{person}{Ruiqi Zhong}, \bibinfo{person}{Tao Yu}, {and} \bibinfo{person}{Dan Klein}.} \bibinfo{year}{2020}\natexlab{}.
\newblock \showarticletitle{{Semantic Evaluation for Text-to-{SQL} with Distilled Test Suites}}. In \bibinfo{booktitle}{\emph{{Proceedings of the 2020 Conference on Empirical Methods in Natural Language Processing (EMNLP)}}}, \bibfield{editor}{\bibinfo{person}{Bonnie Webber}, \bibinfo{person}{Trevor Cohn}, \bibinfo{person}{Yulan He}, {and} \bibinfo{person}{Yang Liu}} (Eds.). \bibinfo{publisher}{Association for Computational Linguistics}, \bibinfo{address}{Online}, \bibinfo{pages}{396--411}.
\newblock
\href{https://doi.org/10.18653/v1/2020.emnlp-main.29}{doi:\nolinkurl{10.18653/v1/2020.emnlp-main.29}}


\end{thebibliography}










\end{document}